\documentclass[runningheads]{llncs}
\usepackage[T1]{fontenc}
\usepackage[colorlinks,allcolors=black,pdftex]{hyperref}
\usepackage{graphicx}
\usepackage{amsmath}
\usepackage{booktabs}
\usepackage{multirow}
\usepackage{algpseudocode}
\usepackage{algorithm}

\begin{document}
\title{Cyclic Data Streaming on GPUs {for Short Range Stencils} Applied to Molecular Dynamics}
\titlerunning{Cyclic Data Streaming on GPUs}
%
\author{Martin Rose\inst{1}\orcidID{0009-0000-9029-2622} \and
Simon Homes\inst{2}\orcidID{0009-0008-3838-0930} \and
Lukas Ramsperger\inst{1}\orcidID{0009-0003-1411-7162} \and
Jose Gracia\inst{1}\orcidID{0000-0002-8925-6592} \and
Christoph Niethammer\inst{1}\orcidID{0000-0002-3840-1016} \and
Jadran Vrabec\inst{2}\orcidID{0000-0002-7947-4051}
}

\authorrunning{M. Rose et al.}
%
\institute{HLRS, University of Stuttgart, Stuttgart \email{rose@hlrs.de}\and
Thermodynamik, Technische Universität Berlin, Berlin}
\maketitle              
\begin{abstract}
In the quest for highest performance in scientific computing, we present a novel framework that relies on high-bandwidth communication between GPUs in a compute cluster.
The framework offers linear scaling of performance for explicit algorithms that is only limited by the size of the dataset and the number of GPUs.
Slices of the dataset propagate in a ring of processes (GPUs) from one GPU, where they are processed, to the next, which results in a parallel-in-time parallelization.
The user of the framework has to write GPU kernels that implement the algorithm and provide slices of the dataset.
Knowledge about the underlying parallelization strategy is not required because the communication between processes is carried out by the framework.
As a case study, molecular dynamics simulation based on the Lennard-Jones potential is implemented to measure the performance for a homogeneous fluid.
Single node performance and strong scaling behavior of this framework is compared to LAMMPS, which is outperformed in the strong scaling case.

\end{abstract}

\section{Introduction}
The traditional approach to parallelizing extensive calculations in science and engineering is the decomposition of the computational domain into subdomains that exchange information across the boundaries of neighboring subdomains.
The performance of applications using this approach is primarily limited by imperfect computational load balance and insufficient overlap of useful computations and communication between subdomains.

Rather than reducing communication between processes to a minimum, the novel approach presented here relies on fast interconnects between GPUs within a compute node and across nodes.
It is embodied in a framework that allows for a straightforward implementation of explicit algorithms without the need to implement a domain decomposition scheme and the corresponding communication pattern.
We name the framework presented here "data streaming framework for explicit algorithms" or DSEA.
It sequentially processes slices of the dataset by a ring of GPUs, where each GPU passes completed slices to the next GPU, resulting in a parallelization-in-time.
As a case study, molecular dynamics (MD) simulation based on the Lennard-Jones potential was implemented to investigate the performance scaling and the influence of inter-node bandwidth on performance.

MD simulations, which treat every molecule individually, are a powerful tool to study structure and dynamics of materials under a wide range of boundary conditions.
Because it rests on a sound physical basis, MD requires few model assumptions, mainly limited to molecular interactions, and the resulting outcomes are of atomistic precision.
From a mathematical point of view, MD can be described as solving $N$ coupled equations of motion, where $N$ is the number of molecules.
Sovling this set of differential equations is a demanding computational task, because the coupling needs to be considered in each timestep.
Next to simulations of homogeneous states, MD is used to investigate transient processes on small scales.
While a few thousand molecules are sufficient to predict simple properties like the pressure of a bulk phase at a given temperature and density, much more molecules are required for the simulation of processes on the nanoscale, like droplet collisions~\cite{Tugend2025} or evaporation~\cite{Homes2021,Homes2023}.
To bridge the gap to larger scales, that can be covered by more approximate methods, like smooth particle hydrodynamics (SPH) or computational fluid dynamics (CFD), it is beneficial to run even larger atomistic MD simulations.

\section{Related Work}

The framework presented in this work belongs to the class of direct space-time parallel solvers in the field of parallel-in-time (PIT) integration methods \cite{tptih5}.
Other solvers in this class have recently been applied to solve ordinary differential equations (ODEs) \cite{Maday2008,Gander2013}.
To our knowledge, there is no PIT method that solves a set of coupled ODEs where the coupling varies over time, as in MD.
Due to the chaotic nature of MD, PIT methods that rely on the prediction of the solution are not considered here.
GStream is a framework that offers data streaming among GPUs \cite{gstream}.
It has been applied to MD, but no improvements of performance or strong scaling have been demonstrated.
GStream does not allow for in-time parallelism or use multi-rail communication between nodes.

\section{Data Streaming Framework}
\label{sec:framework}

In DSEA, the dataset is processed in spatial slices that must be prepared by the user in a preprocessing step.
The Cartesian grid was used in this work, but the concept of slices is not limited to this type of grid.
The slices must be prepared to meet the following requirements:
($1$) All slices are adjoined by one neighbor on the left and one neighbor to the right.
($2$) The spatial shape of a slice can be chosen arbitrarily.
The elements comprised in a slice can be migrated among neighboring slices to balance the computational load as long as condition ($1$) is met.
($3$) Slices must be prepared such that the first and the last slice do not interact, because the framework does not allow for the exchange of information between the first and the last slice.
Requirement ($3$) prevents the implementation of periodic boundary conditions in the direction perpendicular to the slices.
Slices should be defined perpendicular to the longest axis of the dataset to maximize the number of slices ($N_S$) which allows for the use of more GPUs in parallel.
In the application cases described below, this will be addressed as x-axis.


When processing a certain slice $i$, data from neighboring slices might be required, depending on the algorithm.
The output of an explicit algorithm contributes to slice $i$ and to a number of neighboring slices.
The number of neighboring slices accessed as input or output is referred to as \textit{input-order}~($O_\text{in}$) and \textit{output-order}~($O_\text{out}$), respectively.
Any order can be zero, which means that only data from slice $i$ are accessed.

\subsection{Data Flow}\label{sec_data_flow}
\begin{figure}
\centering
\begin{minipage}{.48\textwidth}
\centering
  \includegraphics[width=\textwidth]{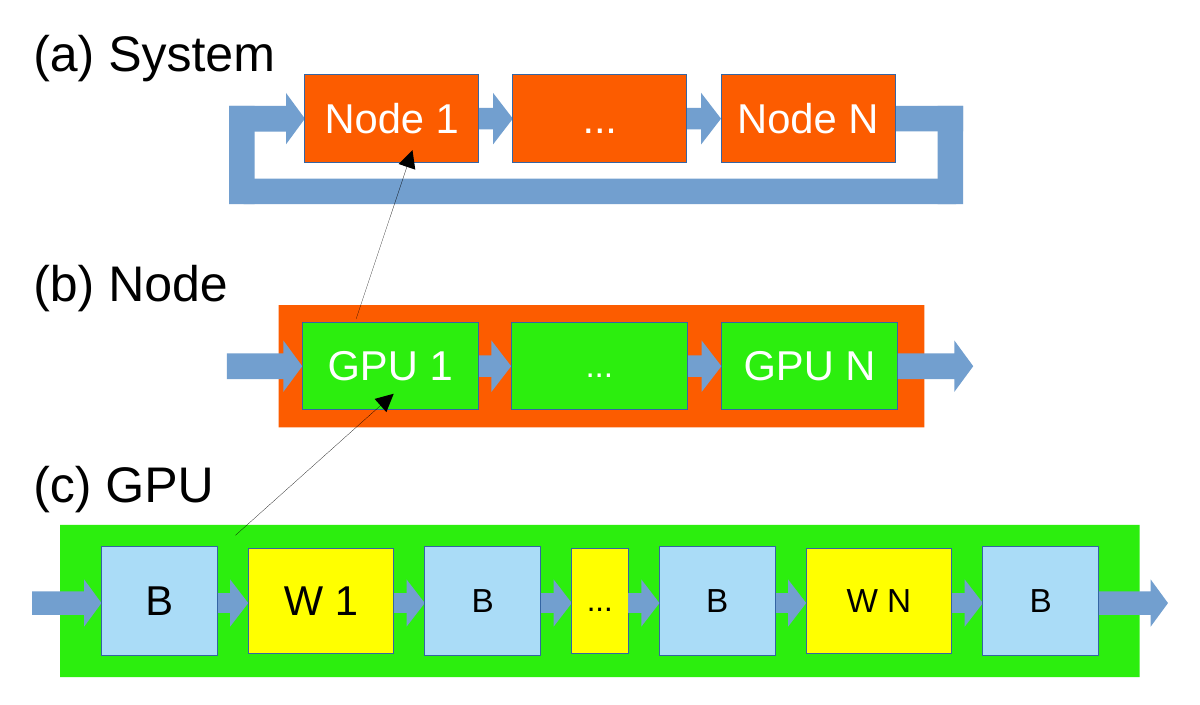}
  \caption{Arrangement of workers (W) and buffers (B). Blue arrows indicate the direction of data flow. (a) System with multiple GPU nodes (red), (b) GPU node with multiple GPUs (green), (c) GPU with multiple workers (yellow).}
  \label{fig_data_flow}
\end{minipage}%
\hfill
\begin{minipage}{.48\textwidth}
\centering
  \includegraphics[width=\textwidth]{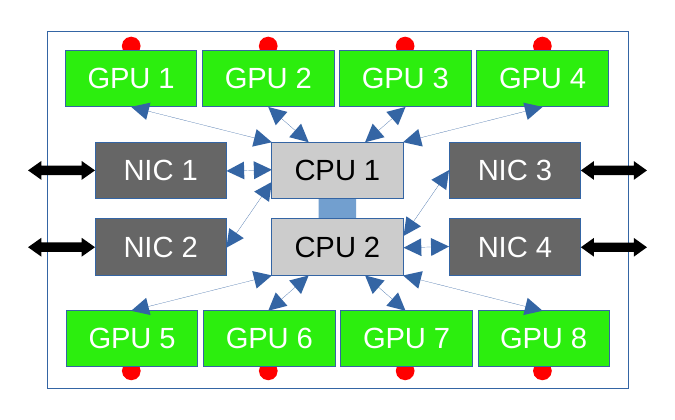}
  \caption{Block diagram of the compute nodes used in this work. Black arrows: 200Gb/s Infiniband, red dots: NVLink connection to NVLink switch, blue arrows: PCIe 4.0 x 16, blue bar: Infinity Fabric.}
  \label{fig:node}
\end{minipage}
\end{figure}

DSEA processes the slices using $N_\text{GPU}$ GPUs by means of $N_\text{wGPU}$ workers per GPU.
The data flow on the system level, the node level and within one GPU is shown in Figure~\ref{fig_data_flow}, where the total number of workers is $N_w = N_\text{GPU} \times N_\text{wGPU}$.
The input and the output of a worker connect to circular buffers that can store multiple slices and reside in GPU memory.
The output of worker $i=1 \dots (N_w-1)$ is the input of worker $i+1$ and the output of worker $i=N_w$ is the input of worker $i=1$.
As a worker advances the timestep for a slice, the dependence between the execution of workers introduces parallelism in time.
It should be pointed out that the $N_w$ workers do not have to cover the entire simulation domain concurrently because slices are stored in buffers before processing.

At the beginning of the computation, slices are read from storage and fed into worker $i=1$.
Once slices have passed through all workers, a \textit{super-cycle} is completed.
In one super-cycle, each of the $N_w$ workers has processed the complete dataset once, which means that the algorithm has advanced $N_w$ time steps.
At the end of the computation, when the last super-cycle is completed, the slices are written to storage.
Loading slices at the beginning of the first super-cycle can be parallelized efficiently because all compute nodes can participate in the I/O operation: All processes load slices in a round-robin fashion and send the slices to the first GPU.
Storage of slices at the end of the last super-cycle can be parallelized analogously.

$N_\text{wGPU}$ workers operate sequentially on a single GPU. The first worker on each GPU connects to an input buffer and the last worker on each GPU connects to an output buffer.
The connection between the GPUs is established between the output buffer of the last worker on GPU $i$ and the input buffer of the first worker on GPU $i+1$.
The output buffer of the last worker on the last GPU connects to the input buffer of the first worker on the first GPU.
The total number of buffers is thus $N_b=N_w+N_\text{GPU}$.
Each of these buffers contains a given number of slots $s$.
In order to store all slices, $s$ must be specified such that $s \times N_b > N_S$.

\subsection {Multi-Rail Inter-Node Data Transfer}
A block diagram of the GPU nodes used in this work is shown in Figure \ref{fig:node}.
The GPUs (NVIDIA~A100, 40~GB) and the NICs (ConnectX-6) were connected to the CPUs (AMD EPYC 7702) via PCIe~4.0~x16.
The GPU nodes formed a fat-tree network.
The bandwidth achieved per NIC was about 20~GB/s for a message size above 1 MB. 
All GPUs were connected via NVLink switches (not shown) with an effective bandwidth of about 250~GB/s between GPUs.
The CPUs were connected via Infinifty Fabric and each CPU had 512~GB of main memory (not shown).
When executing the framework on several of these nodes, the bandwidth between GPUs that are part of different nodes is much lower than the bandwidth between GPUs within a node, if only one NIC is used.
To reduce this bottleneck, a data transfer scheme was implemented that uses up to four NICs simultaneously employing the UCX library \cite{shamis2015ucx}.
To transfer data between nodes, e.g., from GPU~8 of node $k$ to GPU~1 of node $k+1$, the data are first distributed among multiple GPUs via NVLink and then sent from all GPUs via multiple NICs simultaneously.
On node $k+1$, data are received by multiple NICs and stored in multiple GPUs before being combined in GPU $1$ via NVLink.

\subsection{Worker}
The worker performs the computations that realize the algorithm by executing any combination of host functions, GPU kernels and data transfers between host and device.
For a worker to start execution, $2 \times O_{in}+1$ slices must be present in the input buffer.
Moreover, a free slot in the output buffer is required if the worker produces a first contribution to a slice.
The data in the input slices are not overwritten because they are required to process the next slice.
An active worker produces output data for $2 \times O_{out}+1$ slices.
The contribution to a slice finalizes that slice, if the worker will not contribute to it in the next stage.
The user of DSEA has to write only the application-specific kernels for CPU and GPU.

The execution of workers on a single GPU is organized by a state machine that operates in stages until all slices have been received, processed and sent out of a GPU.
In each stage, one slice can be received into the input buffer, another slice can be sent from the output buffer and workers can execute if the required slices are present in their input buffer.

Table \ref{tab_one_worker_stages} illustrates the operation of a single worker for $O_\text{in}=O_\text{out}=1$.
From left to right, the columns of the table correspond to the input buffer, the worker and the output buffer.
The numbers in the columns indicate the slices.

\begin{table}
\centering
\caption{Operation of one worker on a single GPU for $O_{in}=O_{out}=1$.}
\label{tab_one_worker_stages}
\begin{tabular}{c|cc|c|cc}
\multirow{2}{*}{stage} & \multicolumn{2}{c|}{\begin{tabular}[c]{@{}c@{}}buffer\_in\\ input to worker\_a,\\ receive slices\\ $t=t_1$\end{tabular}} & \begin{tabular}[c]{@{}c@{}}worker\_a\\ $t=t_1$\end{tabular} & \multicolumn{2}{c}{\begin{tabular}[c]{@{}c@{}}buffer\_out\\ output of worker\_a,\\ send slices\\ $t=t_2$\end{tabular}} \\
                       & \multicolumn{1}{c|}{store}                                                  & receive                             &                                                          & \multicolumn{1}{c|}{partial}                                         & send                                    \\ \hline
1                      & \multicolumn{1}{c|}{-}                                                      & 1                                   & idle                                                     & \multicolumn{1}{c|}{-}                                               & -                                       \\ \hline
2                      & \multicolumn{1}{c|}{1}                                                      & 2                                   & idle                                                     & \multicolumn{1}{c|}{-}                                               & -                                       \\ \hline
3                      & \multicolumn{1}{c|}{1, 2}                                                   & 3                                   & 1                                                        & \multicolumn{1}{c|}{1, 2}                                            & -                                       \\ \hline
4                      & \multicolumn{1}{c|}{1, 2, 3}                                                & 4                                   & 2                                                        & \multicolumn{1}{c|}{2, 3}                                            & 1                                       \\ \hline
5                      & \multicolumn{1}{c|}{2, 3, 4}                                                & 5                                   & 3                                                        & \multicolumn{1}{c|}{3, 4}                                             & 2                                       \\ \hline
...                    & \multicolumn{1}{c|}{...}                                                    & ...                                 & ...                                                      & \multicolumn{1}{c|}{...}                                             & ...                                     \\ \hline
$N_S$                      & \multicolumn{1}{c|}{$N_S-3$, $N_S-2$, $N_S-1$}                              & $N_S$                               & $N_S-2$                                                  & \multicolumn{1}{c|}{$N_S-2, N_S-1$}                                  & $N_S-3$                                 \\ \hline
$N_S+1$                      & \multicolumn{1}{c|}{$N_S-2$, $N_S-1$, $N_S$}                                & -                                   & $N_S-1$                                                  & \multicolumn{1}{c|}{$N_S-1, N_S$}                                    & $N_S-2$                                 \\ \hline
$N_S+2$                      & \multicolumn{1}{c|}{$N_S-2$, $N_S-1$, $N_S$}                                & -                                   & $N_S$                                                    & \multicolumn{1}{c|}{$N_S$}                                           & $N_S-1$                                 \\ \hline
$N_S+3$                      & \multicolumn{1}{c|}{-}                                                      & -                                   & -                                                        & \multicolumn{1}{c|}{-}                                               & $N_S$                                  
\end{tabular}
\end{table}

The columns representing buffers list the content of that buffer at each stage.
A slice that is stored in a buffer can serve as input for a worker.
Parts of a slice can also be present in a buffer, which means that the worker will contribute more data in the following stage.
Finally, a complete slice can be received or sent from a buffer.
The column that represents the worker lists the central slice that the worker processes at each stage.
The workers are idle if the input buffer does not contain the necessary slices.
By design of DSEA, data transfers to and from buffers as well as the execution of a single worker can occur concurrently.

During the first two stages, the worker is idle while the required slices (one and two) are received by buffer\_in.
Worker\_a starts processing slice one in stage three and contributes data to slices one and two stored in buffer\_out.
No slice is completed in stage three, so no slice can be sent out.
In stage four, worker\_a processes slice two and contributes data to slices one, two and three.
Slice one is now complete and sent out.
In the following stages, slices will be received, processed and sent out until the last slice is received by buffer\_in.
Once worker\_a has processed the last slice $N_S$, it is sent out in the final stage and the super-cycle is completed.

In the fourth stage, slice four is received while slice one is sent out.
This means that four slices are required by the configuration shown in Table~\ref{tab_one_worker_stages} to operate, and $N_{max}=N_S/4$ processes can work in parallel.
In general
\begin{equation}
N_{max}=N_S/(2+N_\text{wGPU}\times(O_\text{in}+O_\text{out})).
\label{eq:nmax}
\end{equation}

\subsection{Program structure}
The DSEA framework is implemented as an MPI application in which one MPI process uses one GPU.
The MPI processes communicate in a ring due to the data flow pattern described in section \ref{sec_data_flow}.
At the start of the program, the status of the slots in all buffers is \textit{free}.
Four OpenMP threads per process are used to realize the following concurrent tasks:
\begin{enumerate}
    \item \textbf{Input}: This thread receives slices from the preceding process in the ring and stores them in the next \textit{free} slot of the input buffer on the GPU.
    Once a slice has been received, the status of the corresponding slot is set to \textit{ready}.
    For the first GPU in the first super-cycle, this thread is responsible for loading slices from storage with the help of the I/O threads of all processes.
    \item \textbf{Output}: This thread sends slices from slots of the output buffer that have reached the state \textit{ready} to the next MPI process in the ring. For the last GPU in the last super-cycle, this thread is responsible for storing slices with the help of the I/O threads of all processes.
    \item \textbf{Main}: The main thread is running the state machine that controls the execution of the workers.
    If a worker is active, this thread executes the required GPU kernels.
    
    \item \textbf{I/O}: This thread performs the I/O operation in the first and last super-cycle.
    When loading slices, process $i$ ($i\geq0$) loads slice $j \times N_{GPU}+i$ for all $j\geq0$ with $j \times N_{GPU}+i<N_S$. After loading, the slice is sent to the first MPI process. An equivalent logic is used for storing slices in the last super-cycle.
\end{enumerate}

\section{Molecular Dynamics in a Data Streaming Fashion}
As a case study, a homogeneous fluid defined by the truncated and shifted Lennard-Jones~(12,6) potential at temperature $T$ and density $\rho$ in a cubic volume of edge length $b$ was considered.
By definition, the molecular interactions were truncated when the distance between a pair of molecules exceeded $r_c$.
The initial state of the simulation was a face-centered cubic (fcc) lattice with four molecules per cell.
The molecular velocities were initialized according to the temperature.
The size of the dataset was controlled by the number of lattice cells ($N_i$) along the edge of the simulation volume.
The edge length of the simulation volume was $b=N_i \times (4/\rho)^{1/3}$ and the number of molecules was $N=4 \times N_i^3$.

To process the dataset described above in DSEA, the simulation volume was divided into $N_{xyz}=b/r_c$ cubic cells in each direction.
Because $N_{xyz}$ has to be an integer, the result of the division was rounded down to the next integer so that the length of the cells is given by $l=b/N_{xyz}$ with $l>r_c$~\cite{cell_list}.
The grid was then sliced perpendicular to the x-axis into $N_{xyz}$ slices, each containing $N_{xyz}^2$ cells.
For each slice, molecules were stored in a continuous list (MolList).
The data structure storing a single molecule contains its position vector $\textbf{r}$, velocity vector $\textbf{v}$ and force vector calculated in the current and the previous time step, $\textbf{F}_{new}$ and $\textbf{F}_{old}$.
All components were stored as 64-bit floating-point numbers.
The indices of molecules in MolList that belong to a given cell were stored in a separate list for each cell (CellList).
The number of molecules in each cell of a slice were stored in CellNM.
MolList, CellList, CellNM and a header containing general information were stored in a continuous memory area that formed a slice of the dataset.

The calculation of the force acting on a molecule only requires access to the molecules in its cell and those in the adjacent cells because $l>r_c$.
In the x-direction, one slice to the left and one slice to the right are required as input to provide the molecules in neighboring cells.
When a molecule propagates, it can reach a cell in the same slice, a cell in the slice to the left or a cell in the slice to the right.
Therefore, $O_{in}=O_{out}=1$ holds for molecular dynamics with the Lennard-Jones potential in this framework because the width of the slices was longer than the cutoff radius up to which the molecular interactions were explicitly evaluated.

The computations described below were conducted with the truncated and shifted Lennard-Jones potential, i.e., $\sigma=1$ and $\varepsilon=1$, and a cutoff radius of $r_c=2.5\,\sigma$.
The timestep was specified to be $\Delta t \sqrt{\varepsilon/m}/\sigma \approx 1.8\cdot10^{-3}$.

\begin{algorithm}[ht]
\caption{Velocity-Verlet algorithm used in DSEAmd with the number of particles $N$, force $\textbf{F}$, position/distance $\textbf{r}$, particle velocity $\textbf{v}$, potential energy $U$ and virial $V$ (for pressure calculation).}

\label{alg_md}
\begin{algorithmic}

\Repeat
    \For{$i=1$ to $N$}
        \State $\textbf{F}_{i old}=\textbf{F}_{i new}$
        \State $\textbf{F}_{i new}=0$
        \For{$j=1$ to $M(i)$} \Comment{loop over $M(i)$ neighbors}
            \State $\textbf{r}_{ij}=\textbf{r}_{i}-\textbf{r}_{j}$
            \If{$\textbf{r}_{ij}*\textbf{r}_{ij}<=r_c^2$} \Comment{molecule j within cutoff radius}
                \State $F_{abs}=24*(2\sigma^{12}/r_{ij}^{12}-\sigma^{6}/r_{ij}^6)/r_{ij}^{2}$
                \State $\textbf{F}_{i new}=\textbf{F}_{i new}+\textbf{r}_{ij}*F_{abs}$ \Comment{accumulate force}
                \State $U=U+4*(\sigma^{12}/r_{ij}^{12} - \sigma^{6}/r_{ij}^{6} + U_{shift})/2$ \Comment{accum. pot. energy}
                \State $V=V+(2\sigma^{12}/r_{ij}^{12} - \sigma^{6}/r_{ij}^{6})/2$ \Comment{accumulate virial}
            \EndIf
        \EndFor
    \EndFor

    \For{$i=1$ to $N$}
        \State $\textbf{v}_\textbf{i}=\textbf{v}_{i}+(\textbf{F}_{i new}+\textbf{F}_{i old})*0.5*dt$ \Comment{update velocity of molecule $i$}
    \EndFor

    \For{$i=1$ to $N$}
        \State $\textbf{r}_{i}=\textbf{r}_{i}+\textbf{v}_{i}*dt+\textbf{F}_{i new}*0.5*dt^2;$ \Comment{update position of molecule $i$}
    \EndFor

\Until{simulation completed}

\end{algorithmic}
\end{algorithm}

\subsection{Implementation of GPU Kernels}
We refer to the present implementation of MD in DSEA as DSEAmd.
Listing \ref{alg_md} shows the Velocity-Verlet algorithm \cite{Allen1987}.
The first loop computes the new force acting on each molecule as a result of the interactions with the molecules within the cutoff radius.
The second loop calculates new velocities from the resulting forces and the forces of the previous time step.
The third loop calculates new positions of all particles from the velocities and the acting forces.

These loops were implemented as a set of GPU kernels that are described below.
The description of the GPU kernels focuses on the details that lead to high compute performance, while we refer the reader to the source code \cite{dsea_source_code} to achieve a complete understanding of the implementation.
Text in italics refers to function names and GPU kernels that are called in the host function \textit{caller\_worker}.
The kernels that perform computations for all molecules, process one molecule per thread.
The index of the molecule to process is obtained from a list of molecule indices built by the kernel \textit{md\_fill\_mol\_work\_list}.
The indices are ordered by cell, starting with the first molecule in the first cell.
This order of molecules leads to neighboring threads processing molecules in the same cell or in a cell close by.
The kernel \textit{md\_v3a} computes the new force acting on each molecule and needs to read the corresponding molecules from neighboring cells.
Therefore, this kernel takes the longest time to complete.
The performance of this kernel depends significantly on the order in which molecules are processed.
Different threads that process molecules in the same cell need to read the same list of neighboring molecules, while threads that process molecules from different cells need to read different lists of molecules.
The extreme case with the lowest performance is when every thread has to read a different list of molecules.
This scenario occurs when molecules from many different cells are processed by a group of 32 GPU threads working in lockstep.
Another factor determining the performance of this kernel is the maximum number of neighbors $M$ of a single molecule in a group of 32 molecules.
$M$ increases drastically when a group of 32 molecules contains molecules from low-density and high-density regions, e.g. at phase boundaries.
New velocities are calculated for all molecules by kernel \textit{md\_v3aa}.
Kernels \textit{md\_thermo\_a} and \textit{md\_thermo\_b} compute the scale factor for the velocities required by the thermostat to keep the temperature of the fluid constant.
Kernel \textit{stat\_collect} collects the contribution from the current cycle to the potential energy $U$ and the virial $V$.
Kernel \textit{md\_v3b} applies the thermostat by scaling the molecular velocities.
Moreover, new positions are computed and the molecules are stored.
In this kernel, the molecules might migrate to a neighboring slice so that a decision about their storage location must be made at runtime, once the new coordinate along the x-axis was calculated.

\subsection{Validation}
\label{sec:validation}
\begin{figure}
  \centering
  \includegraphics[width=0.9\textwidth]{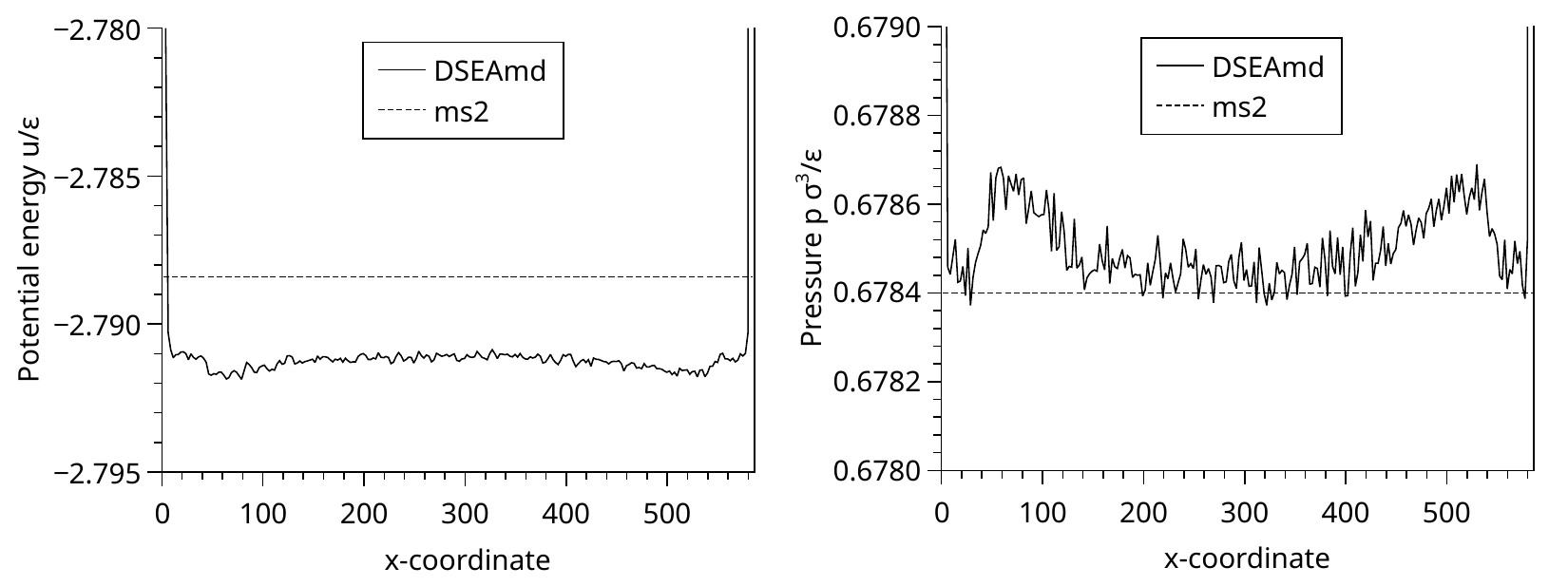}
  \caption{Potential energy $u$ and pressure $p$ obtained with DSEAmd in comparison to results obtained with \emph{ms2}.}
  \label{fig_validate}
\end{figure}

DSEAmd was validated by simulating a bulk liquid consisting of $N=10^8$ molecules at a temperature of $k_{\mathrm B}T/\varepsilon=1.5$ and density $\rho\sigma^3=0.5$, which is a dense supercritical state characterized by rapid molecular propagation and numerous interactions leading to a high pressure.
The results for the most important thermodynamic properties were compared to those obtained with the well-established MD code \emph{ms2}~\cite{Fingerhut2021} for a system size of $N=2048$ molecules.
While DSEAmd yields spatially resolved results, \emph{ms2} only gives values averaged over the entire simulation domain.
Since the validation simulations were executed in the frequently used canonical ($NVT$) ensemble, in which the number of particles $N$, volume $V$ and temperature $T$ are kept constant, the potential energy $u$ and pressure $p$ were used for comparison, cf. Figure~\ref{fig_validate}.
It can be seen that the potential energy values calculated with \emph{ms2} and DSEAmd agree well with a deviation of about $0.1\%$.
This difference can be explained by the finite-size effect due to the very different number of molecules $N$.
For the pressure, both codes yield an almost identical result, differing by only about $0.01\%$.
Larger deviations are observed only close to the domain boundaries, where a mirror is present in DSEAmd, while ms2 operates with periodic boundary conditions in all directions.
This leads to a slight increase in the pressure profile of about $0.03\%$.
It can be concluded that DSEAmd yields results that are confirmed by \emph{ms2}.

\subsection{Performance Scaling}
\label{sec:performance}
The computational performance in DSEAmd is defined as the ratio between the total number of molecules processed by all workers in one super-cycle divided by the time it takes for all slices to complete one super-cycle.
The performance of DSEAmd was compared to that of LAMMPS~\cite{Thompson2022} for strong scaling scenarios.

\begin{figure}
\centering
\begin{minipage}{.48\textwidth}
\centering
  \includegraphics[width=\textwidth]{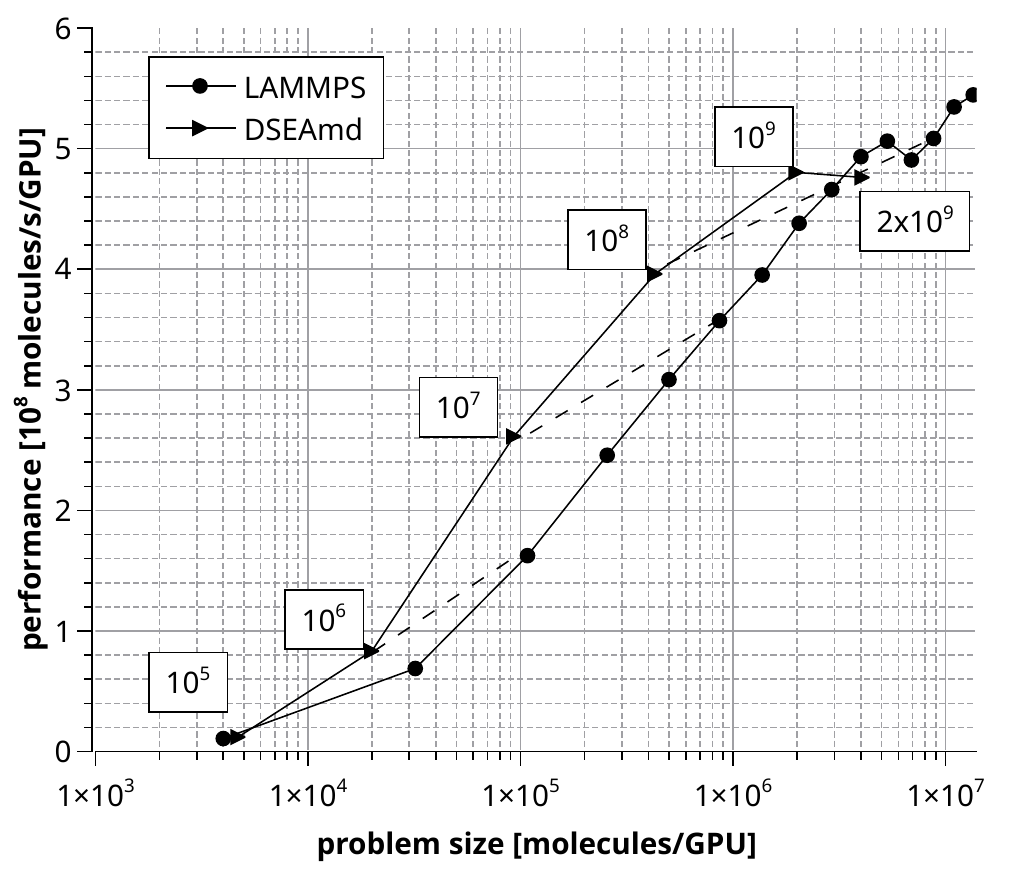}
  \caption{Performance as a function of problem size. Boxes indicate total number of molecules for DSEAmd. Dashed lines connect to corresponding LAMMPS cases.}
  \label{fig_perf_vs_size}
\end{minipage}%
\hfill
\begin{minipage}{.48\textwidth}
\centering
  \includegraphics[width=\textwidth]{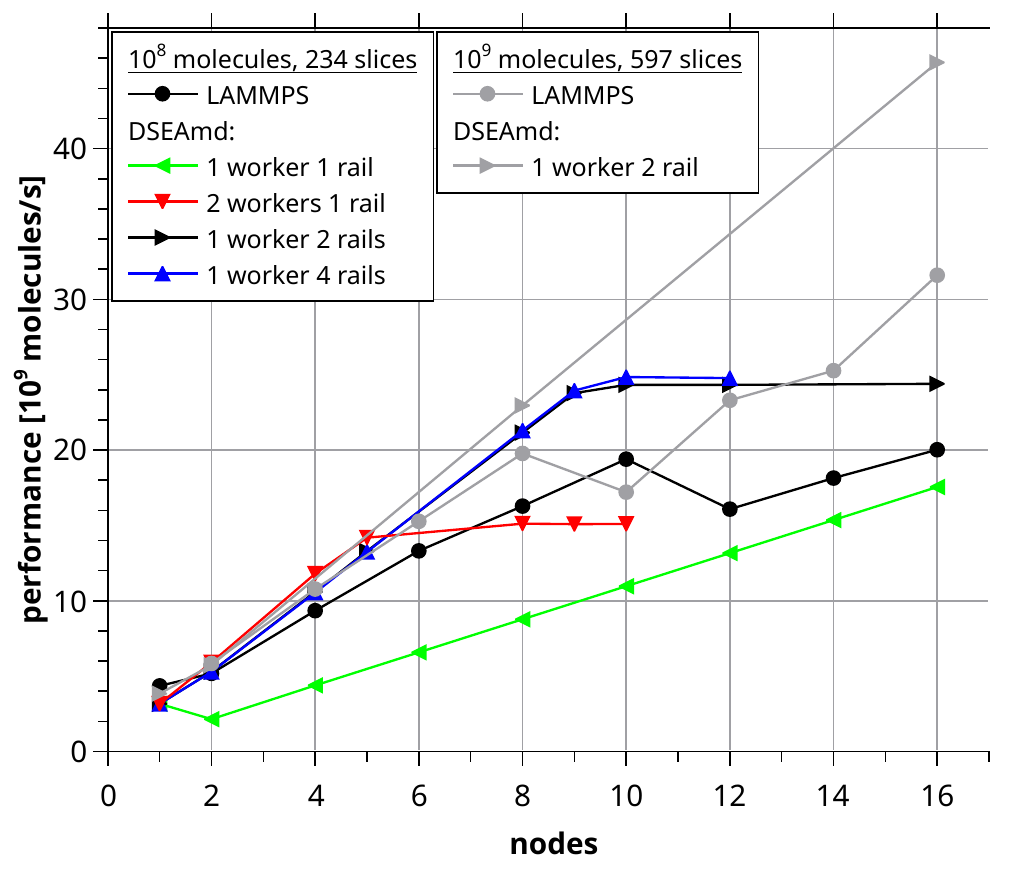}
  \caption{Performance of DSEAmd and LAMMPS as a function of number of nodes, workers per GPU and rails for $10^8$ and $10^9$ molecules.}
  \label{fig_perf_scaling_293_630}
\end{minipage}
\end{figure}

Figure \ref{fig_perf_vs_size} shows the performance obtained on a single compute node using all eight GPUs for DSEAmd and LAMMPS.
In DSEAmd, the problem size per GPU is determined by the number of cells in a slice because a worker in DSEA processes only one slice at a time.
As LAMMPS uses a strategy of spatial domain decomposition, the problem size per GPU is the total number of molecules divided by the number of GPUs.
The dashed lines in Figure \ref{fig_perf_vs_size} connect datasets with the same total number of molecules.
Due to the larger problem size per GPU for a given total number of molecules, LAMMPS outperforms DSEAmd in the single node benchmark.

We investigated the strong scaling of the performance by increasing the number of GPU nodes.
Figure \ref{fig_perf_scaling_293_630} shows the measured performance of DSEAmd and LAMMPS for 1 to 16 nodes and $10^8$ molecules.
For LAMMPS, we observed a speedup of $1.19$ when using two instead of one node and a speedup of $1.8$ when using four instead of two nodes.
This means that the scaling of performance is significantly reduced by communication across node boundaries.
The highest performance of $2 \times 10^{10}$ molecules/s was observed for 16 nodes, while a fluctuation of performance scaling is present between 10 and 16 nodes.
In DSEA, the transfer of a slice between GPUs takes place while the GPUs are computing.
If the transfer time is longer than the compute time, a bottleneck is introduced.
The performance of DSEAmd was measured for a varying numbers of rails for communication and varying numbers of workers per GPU.
For one worker and one rail, a significantly reduced performance was observed when using two instead of one node.
Starting from two nodes, a linear increase of performance that is expected from the design of DSEA was confirmed.
Using two workers and one rail, the performance scales linearly from one to five nodes.
Beyond that, performance does not increase further because there are insufficient slices to serve as input for all workers on all GPUs.
The bottleneck observed for two nodes with one worker and one rail was removed by extending the compute time by a factor of two, i.e. two workers operate sequentially on one GPU at the cost of using more slices on one GPU. 
Using one worker and two rails, a speedup of 1.67 was achieved when using two nodes instead of one.
For two to nine nodes, the scaling of performance is linear, and for ten or more nodes, the performance is constant.
With the combination of one worker and two rails, DSEAmd outperformed LAMMPS.
The bottleneck observed for two nodes was significantly reduced, but not removed.
It did not recede when using four rails instead of two, while the performance at the plateau was slightly higher.

The performance of LAMMPS and DSEAmd using one worker and two rails for the largest dataset investigated in this work with $10^9$ molecules is shown in Figure \ref{fig_perf_scaling_293_630}.
The speedup of DSEAmd from one to two nodes was 1.49 and performance scaled linearly from 2 to 16 nodes.
DSEAmd clearly outperformed LAMMPS for the range of nodes investigated.
Linear scaling of DSEAmd is expected for up to 149 GPUs using Equation (\ref{eq:nmax}) with an extrapolated performance of about $5.1\times10^{10}$ molecules/s.

\section{Summary and Outlook}
We presented a new framework called DSEA, which allows for the implementation of explicit algorithms for the execution on GPU clusters.
The user of this framework has to write GPU kernels that implement the algorithm, while the framework itself provides highly efficient parallelization.
The dataset is partitioned into slices that are processed by a ring of processes, which results in a parallelization-in-time.
Molecular dynamics simulation (DSEAmd) was implemented in DSEA to investigate the scaling of the computational performance.
A linear scaling of performance was observed together with the expected plateau in performance.
In comparison to LAMMPS, the performance of DSEAmd is higher in strong scaling benchmarks for the case investigated here.
To our knowledge, this is the first application of parallel in-time integration to a set of coupled differential equations with a time depend coupling.

In the next step, the effects of computational load balance on performance of DSEAmd will be studied.
We will implement direct numerical simulation in DSEA and compare performance to other codes.

\begin{credits}
\subsubsection{\ackname}
The authors gratefully acknowledge the financial support by the Federal Ministry of Education and Research (BMBF) under the grant WindHPC (grant number 16ME0608K).
This work has received funding in the context of the ChEESE-2P project from the European High Performance Computing Joint Undertaking (JU) under grant agreement No. 101093038 and the German Federal Ministry of Education and Research, grant No. 16HPC081.

\subsubsection{\discintname}
The authors have no competing interests to declare that are relevant to the content of this article.
\end{credits}
%
%
%
\bibliographystyle{splncs04}
\bibliography{dsea_lncs.bib}

\end{document}